\documentclass[12pt]{article}
\usepackage[top=1in, bottom=1in, left=1.in, right=1.in]{geometry}
\usepackage[english]{babel}
\usepackage{atbegshi,cite}
\usepackage{amsmath,amssymb,amsbsy,amstext, amsthm, simplewick}
\usepackage{hyperref}
\usepackage{graphicx}
\usepackage{amsfonts}
\usepackage[small]{caption}
\usepackage{upgreek}
\usepackage[titletoc]{appendix}
\usepackage{setspace}
\usepackage[dvipsnames]{xcolor}
\usepackage{slashed}
\usepackage{comment}
\usepackage{soul}

\newcommand{\newc}{\newcommand}

\newc{\sflip}{{\bf \color{red} ***sign***}}
\newc{\TR}{{\rm Tr}}
\newc{\Zbb}{{\mathbb Z}}
\newc{\Rt}{{\mathbb R}^3}
\newc{\Rtw}{{\mathbb R}^2}
\newc{\Rf}{{\mathbb R}^4}
\newc{\So}{{\mathbb S}^1}
\newc{\Stw}{{\mathbb S}^2}
\newc{\Sth}{{\mathbb S}^3}
\newc{\zt}{{\mathbb Z}_2}
\newc{\RtSo}{{\mathbb R}^3\times{\mathbb S}^1}
\newc{\RtwStw}{{\mathbb R}^2\times{\mathbb S}^2}
\newc{\scriminus}{{\cal I}^-}
\newc{\scriplus}{{\cal I}^+}
\newc{\mpl}{M_p}
\newc{\Ricci}{\mathcal{R}}
\newc{\calU}{{\cal U}}
\newc{\calK}{K}
\newc{\calUi}{{\cal U}^{-1}}
\newc{\calG}{{\cal G}}
\newc{\calM}{{\cal M}}
\newc{\calL}{{\cal L}}
\newc{\calO}{{\cal O}}
\newc{\calR}{{\cal R}}
\newc{\calQ}{{\cal Q}}
\newc{\calI}{{\cal I}}
\newc{\calOb}{{\cal O}^\dagger}
\newc{\IEH}{I_{\calL}}
\newc{\Irad}{I_{\rm ADM}^{\rm radial}}
\newc{\Itemp}{I_{\rm ADM}^{\rm temporal}}

\begin{document}
\begin{titlepage}
\begin{flushright}
{\large 
~\\
}
\end{flushright}

\vskip 2.2cm

\begin{center}

{\large \bf Euclidean de Sitter Black Holes and Microcanonical Equilibrium}

\vskip 1.4cm

{ Patrick Draper$^{(a)}$ and Szilard Farkas}
\\
\vskip 1cm
{$^{(a)}$ Department of Physics, University of Illinois, Urbana, IL 61801}\\
\vspace{0.3cm}
\vskip 4pt

\vskip 1.5cm

\begin{abstract}
Schwarzschild-de Sitter (SdS) black holes do not admit a completely smooth Euclidean continuation. We discuss some modifications of the gravitational path integral that give Euclidean SdS a semiclassical equilibrium interpretation. First we consider ``gravity in a cavity," defining the canonical ensemble in a box that excises one horizon. However, this standard approach does not work for positive cosmological constant: the solution of lowest free energy has a negative heat capacity, which is inconsistent if it is to provide the leading semiclassical contribution to a canonical partition function. Instead we modify the boundary conditions in the path integral to construct the microcanonical partition function, which appears to be well-defined. We then bring two ensembles into contact and remove the boundary, producing states of a larger microcanonical ensemble that contain, for example, both a black hole and a cosmological horizon at once. These systems are closed and have no boundary, but they must possess some form of mild metric discontinuity. We discuss the case where the discontinuity is equivalent to the insertion of a thin, rigid membrane, separating two systems that can exchange energy and are at local equilibrium. Equilibrium configurations obtained in this way are found to be thermodynamically unstable if they contain a black hole.

\end{abstract}

\end{center}

\vskip 1.0 cm

\end{titlepage}
\setcounter{footnote}{0} 
\setcounter{page}{1}
\setcounter{section}{0} \setcounter{subsection}{0}
\setcounter{subsubsection}{0}
\setcounter{figure}{0}

\section{Introduction}
The Schwarzschild-de Sitter (SdS) solutions of Einstein gravity describe spacetimes with two horizons. The first laws can be written as
\begin{align}
    dM&=T_b dS_b\nonumber\\
    dM&=-T_c dS_c
    \label{SdSfirstlaw}
\end{align}
where subscripts refer to the black hole and cosmological horizons, respectively, and both temperatures are positive. The unusual minus sign in the first law for the cosmological horizon means that the entropy of this horizon decreases as the black hole mass increases. Because the total entropy $S_{tot}=S_b+S_c$ is finite and $T_b>T_c$, $S_{tot}$ also decreases with $M$. These observations have been interpreted as implying that (1) pure de Sitter (dS) describes the  equilibrium state of a finite number of degrees of freedom~\cite{Banks:2000fe}, and (2) local excitations like black holes are constrained states of these degrees of freedom, realized in a matrix quantum mechanics model~\cite{Banks:2006rx,Banks:2013fr,Banks:2020zcr,Susskind:2021dfc}). Further investigation of SdS thermodynamics might prove useful to extract the properties of this conjectured matrix model.

Ordinarily, Euclidean continuation is a powerful technique for studying the thermodynamic properties of horizons, and for computing the nonperturbative nucleation rate of black holes at finite temperature. However, the presence of two temperatures in Eq.~(\ref{SdSfirstlaw}) complicates the method for de Sitter black holes. 
A priori, Euclidean SdS has conical singularities at each horizon. One or the other can be removed by choice of thermal periodicity, but then the other cannot be removed if $T_b>T_c$. As a result Euclidean SdS is not a gravitational instanton (except in the special case where the horizon radii coincide, $T_b=T_c$, and the solution is smooth~\cite{Ginsparg:1982rs}.) In this work we will examine some methods by which Euclidean continuation can still be used to probe the thermodynamics of dS black holes. Our primary methods involve placing the system in a cavity; adjusting the boundary conditions on the cavity walls (which amounts to changing the thermodynamic ensemble); and recombining cavities to produce closed systems in and out of thermal equilibrium.

If the thermal periodicity in SdS is chosen to remove one conical singularity, a useful way to deal with the other is to excise it and add a boundary. The result is a genuine stationary point of the action. We can also move this boundary around, and adjust the local temperature on it. We consider only spherically symmetric solutions in this work and the boundary is always prescribed at a fixed radius. Then  the solutions with boundaries fall into two different classes, depending on which conical singularity -- $r_b$ or $r_c$ -- is excised. If the boundary eliminates the cosmological horizon, there are one, two, or three solutions consistent with the  given induced metric on the boundary. We refer to these as ``black hole side" solutions. If the boundary eliminates the black hole horizon, there is zero or one solution, which we refer to as the ``cosmological side" solution. The thermodynamics of these solutions are discussed in Sec.~\ref{sec:SdScavities}.

The black hole side solutions have both flat space and AdS analogs. In infinite flat space, the thermal path integral has two saddle point solutions, $R^3\times S^1$ and the Euclidean Schwarzschild geometry. The latter was interpreted by Gross, Perry, and Yaffe (GPY) as an instanton describing the decay of the homogeneous solution by thermal nucleation of black hole states~\cite{Gross:1982cv}. This result was further explained in an analysis by York, in which the problem was analyzed at finite thermodynamic volume with fixed spatial boundary conditions~\cite{York:1986it}. York found that at high temperatures there are two black hole solutions of different mass that satisfy the boundary conditions, one of which corresponds to the GPY instanton, and the other of which gives the semiclassical approximation to the ground state. The former has a perturbative instability, necessary for its interpretation as a decay event, while the latter has positive heat capacity and is perturbatively stable, which is necessary in order to have a sensible semiclassical approximation to the path integral. The phase transition between the ``hot flat space" and the ``large black hole" solutions can be thought of as a flat-space finite-volume analog to the Hawking-Page transition in AdS. At positive cosmological constant, with the cosmological horizon removed by a boundary, the one, two, or three solutions referred to above play a qualitatively similar role.

The cosmological side solution, on the other hand, is a novel feature of the positive cosmological constant and does not have a flat space or AdS analog. Moreover, it has the lowest free energy of the solutions, so it ought to be regarded as the leading contribution to the partition function with fixed temperature boundary conditions. However, this interpretation is not consistent. The solution possesses a negative heat capacity $C=\beta^2\partial_\beta^2\log Z < 0$, and so cannot be interpreted as the semiclassical approximation to a stable equilibrium state at fixed temperature.

The simplest way to deal with this situation is to fix the boundary energy rather than the temperature. This amounts to a change of boundary conditions in the gravitational path integral, discussed in Sec.~\ref{sec:micro}. In a simple gauge which can always be reached, it is sufficient to fix an extrinsic curvature on the spatial boundary and allow the lapse function to vary. The ensemble becomes microcanonical and the negative heat capacity is no longer inconsistent.  At the leading semiclassical order we find that there is a unique solution at each boundary radius and energy. The relevant solution governs the microcanonical thermodynamic equilibrium. 

It is then also natural to recombine two solutions of the same cavity radius. This is particularly interesting because it allows us to remove the boundaries and consider closed systems without boundaries, closest to the original problem of interest. 
In Sec.~\ref{sec:equilibrium} we discuss equilibrium configurations obtained in this way. The local temperature is continuous, but is a discontinuity in the radial components of the metric that is physically equivalent to the inserting a rigid, massive, heat conducting shell and taking the thin-shell limit. We find that these equilibrium states have the lowest entropy that can be attained in the presence of a black hole, while the highest entropy state with the same total energy corresponds to  empty patch of de Sitter joined by a thin shell to a cosmological side solution. Thus, while the black hole horizon can be put into equilibrium with a de Sitter horizon by the introduction of the membrane, it is not possible to stabilize the black hole state in this way.

This paper is a companion to~\cite{DFeuclidean2}, in which we discuss the problem of computing the black hole nucleation rate out of the empty de Sitter bath using Euclidean methods. In this case there is no membrane or thermodynamic equilibrium between the horizons, but we find that Euclidean SdS provides a genuine stationary point to a constrained path integral. 

As this work was being completed, Ref.~\cite{svesko2022quasilocal} appeared, which examines a related set of questions in 1+1 dimensional JT gravity to those considered in 3+1 dimensional Einstein gravity in this paper.

\section{Solutions with boundaries}
\label{sec:SdScavities}
\subsection{SdS }
First, we briefly review the properties of Euclidean Schwarzschild-de Sitter.  The partition function and Euclidean action are
\begin{align}
&~~~~~~~~~~~~~~~~~~~~Z=\int Dg\, e^{-I[g]}\nonumber\\
I &= -\frac{1}{16\pi}\int_{\calM} d^4x \sqrt{g} \left(\calR-2\Lambda\right)-\frac{1}{8\pi}\int_{\partial\calM} d^3x \sqrt{\gamma} (K-K_0). 
\label{eq:action}
\end{align}
The action is a sum of terms, $I\equiv I_{tot} = I_{EH}+I_{GHY}-I_{sub}$, where $I_{EH}$ and $I_{GHY}$ are the bulk Einstein-Hilbert term and the Gibbons-Hawking-York boundary term, and $I_{sub}$ is the subtraction term $\propto K_0$, which we take to  be the trace of the extrinsic curvature of the boundary surface embedded in flat space. The subtraction term has a certain arbitrariness to it; for our purposes, it is chosen to make a direct connection to the formulas known for a cavity at zero cosmological constant, which are recovered from ours in the  limit that the cosmological constant $\Lambda$ goes to zero.

The Euclidean SdS metric in static coordinates is
\begin{align}
dl^2&= f(\rho) dt^2 + f(\rho)^{-1} d\rho^2+\rho^2 d\Omega^2\nonumber\\
f(\rho)&=1-2M/\rho-(\rho/L)^2.
\label{eq:sdsmetric}
\end{align}
$\rho=r_{b,c}$ are the black hole and cosmological horizon radii and are related to the mass and de Sitter radius $L=\sqrt{3/\Lambda}$ by
\begin{align}
M&=\frac{1}{2}r_b\left(1-\frac{r_b^2}{L^2}\right)=\frac{1}{2}r_c\left(1-\frac{r_c^2}{L^2}\right)\nonumber\\
L^2&=r_b^2+r_br_c+r_c^2.
\label{eq:MLeq}
\end{align}
The quadratic equation relating $r_{b,c}$ and $L$ can also be rearranged as
\begin{align}
    r_{c,b}=\frac{1}{2}\left(\sqrt{4L^2-3r_{b,c}^2}-r_{b,c}\right).
\end{align}

\noindent The horizon temperatures that are defined by surface gravities $T=|f'(r_{b,c})|/4\pi$ 
are
\begin{align}
    T_b=\frac{L^2-3r_b^2}{4\pi L^2 r_b},\;\;\;\;\;\;T_c=\frac{3r_c^2-L^2}{4\pi L^2 r_c},
\end{align}
and the local temperature at $r$ is $T(r)=T_{hor}/\sqrt{f(r)}$. If the Euclidean static time coordinate is taken to have periodicity $t\sim t+1/T_{b(c)}$, there is a conical singularity at $r_{c(b)}$.

Now we consider a sphere $\Sigma$ of radius $\rho=r$. $r$, $L$, and the local temperature at $r$ determine the induced metric on $\Sigma$. We can solve the equations of motion with these boundary conditions to obtain candidate saddle point approximations to the partition function. As described above we write the total action as
\begin{align}
    -\log Z = I_{tot}=I_{EH}+I_{GHY}-I_{sub}.
\end{align}
Then we can vary the local temperature with fixed $r$ to compute other thermodynamic quantities from $\log Z$.  This is equivalent to changing the induced metric on $\Sigma$ so that it takes the form of the $n$-fold replica of the original metric. Therefore this class of variations can be thought of as varying the temperature of a system with fixed Hamiltonian.

We use the boundary to eliminate whichever horizon has the conical singularity.
 Then for a fixed spherically symmetric boundary 3-metric, with boundary specified by static coordinate radius $r$, and dS radius $L$, there are one to four smooth solutions. 
The four solutions correspond to an empty patch of dS with cosmological horizon removed; two black hole side solutions with different $r_b$ and with cosmological horizon removed; and a cosmological side solution where the boundary removes the black hole horizon. We discuss each in turn.

\subsection{Empty dS solution}
The ``empty patch of dS" solution corresponds to taking a region of Euclidean dS with boundary radius $r<L$ and adjusting the strength of the conical singularity at the cosmological horizon so that the temperature at $r$ is $T$. Ordinarily we would choose to remove the conical singularity by  setting the  periodicity of the static time coordinate to $t\sim t+1/T_{dS}$ with $T_{dS}=1/2\pi L$; instead, we choose $T_{dS}=T\sqrt{1-r^2/L^2}$. Since the singularity lies outside the patch of interest, the solution is smooth. The action is
\begin{align}
    I_{EH}+I_{GHY}&=-\frac{1}{2T_{dS}}(r^3/L^2)+\frac{1}{2T_{dS}}(3r^3/L^2-2r)\nonumber\\
    I_{sub}&=-\frac{r}{T_{dS}}\sqrt{1-r^2/L^2}\nonumber\\
    I_{tot}&=I_{EH}+I_{GHY}-I_{sub}.
\end{align}
The total action is nonnegative, and vanishes at $r/L=0,1$. At high temperatures it goes to zero.

Identifying $I_{tot}=-\beta F$, with $\beta= 1/T$, we  compute the thermodynamic energy $E=\partial_\beta (\beta F)$  and entropy $S=-(1-\beta \partial_\beta)\beta F$:
\begin{align}
    &E=r(1-\sqrt{1-r^2/L^2})\nonumber\\
    &S=0.
\end{align}
The energy also satisfies
\begin{align}
    E-\frac{1}{2}\frac{E^2}{r}=\frac{r^3}{2L^2}.
\end{align}
For $r\rightarrow L$  the total thermodynamic energy is $E=L$.

~\\
\subsection{Two BH solutions}
There may also be SdS solutions that satisfy the boundary condition $T(r)$. For a smooth black hole horizon, the static time coordinate is periodic, $t\sim t+1/T_b$.  Then the local temperature at $r$ is $T=T_b/\sqrt{f(r)}$, or
\begin{align}
    T=\frac{L^2-3r_b^2}{4\pi L^2r_b}\frac{1}{\sqrt{1-\frac{r_b}{r}-\frac{r^2}{L^2}+\frac{r_b^3}{r L^2}}}
    \label{eq:rbconstraint}
\end{align}
Regarding this as an equation for $r_b$, physically relevant solutions must have  $0\leq r_b\leq r$. At the ends of this interval, the temperature diverges, and it is bounded away from zero in the interval. There are zero, one, or two solutions.

\begin{figure}[t!]
\begin{center}
\includegraphics[width=0.45\linewidth]{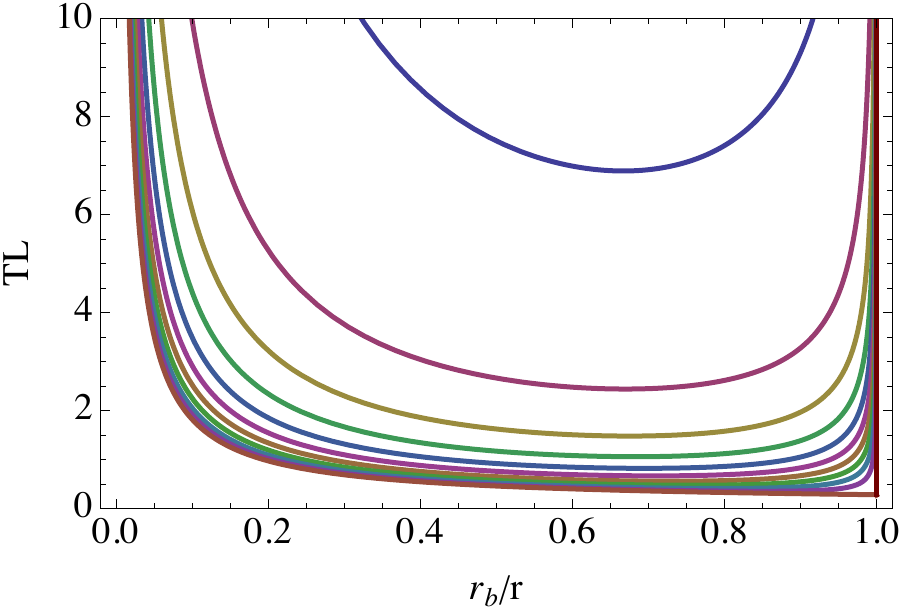}~~~~~~~~
\includegraphics[width=0.45\linewidth]{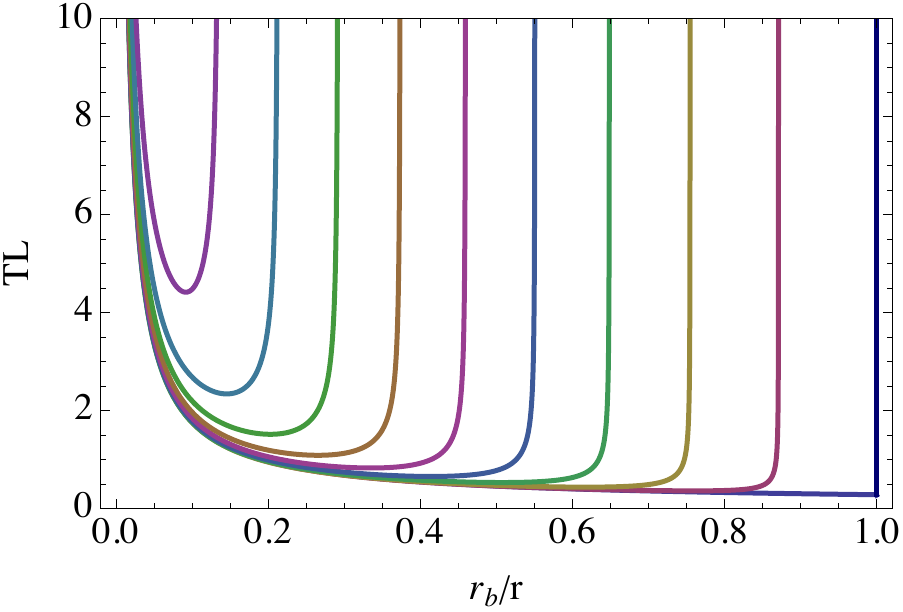}
\caption{Left: Local temperature at boundary radius $r$ in units of the dS curvature scale, for a set of $r$ less than the Nariai radius, as a function of black hole radius and assuming a smooth black hole horizon. Values of $r$ increase from the top curve to the bottom curve ($r=r_N$). Typically there are zero or two solutions for a given $T$, corresponding to a ``small" and a ``large" black hole. When $r=r_N$, there is a second solution for all $T\geq 1/(2\pi r_N)$. Right: the same, but for $r\geq r_N$. Here values of $r$ increase from the bottom curve ($r=r_N$) to the top curve. Again there are two solutions at high temperatures corresponding to a smaller and a larger black hole.}
\label{fig:rbT}
\end{center}
\end{figure}

The temperature as a function of $r_b$ is shown in  Fig.~\ref{fig:rbT}. 
 For $r$ less than the Nariai radius $r_N=L/\sqrt{3}$, at sufficiently large $T$ there are two solutions, $r_b\approx 0$ and $r_b\approx r$. For $r$ greater than the Nariai radius and large $T$, the solutions are approximately $r_b\approx 0$ and the solution to $r_c\approx r$, where $r_c$ and $r_b$ are related by Eq.~(\ref{eq:MLeq}). For generic $r$ there is a minimum temperature where a solution exists and there is only one solution at that temperature; at lower temperatures, there is no solution with a black hole. The minimum temperature for which black hole solutions exist reaches its absolute minimum for $r=r_N$.  The minimum value is $T_{min}=\sqrt{3}/(2\pi L)=1/(2\pi r_N)$ and corresponds to the limit  $r_b\rightarrow r_N$; this is the one case where the temperature does not diverge as $r_{b,c}\rightarrow r$.  A separate analysis is required at $r=r_N$ to see that there is in fact a second solution for $T>T_{min}$.\footnote{To see it one can set $r_b=r_N-\epsilon$ and $r=r_N+a\epsilon$ and take the limit $\epsilon\rightarrow 0$; in this limit the local temperature is $T_{min}/\sqrt{1-a^2}$. This shows that the  Nariai solution can achieve any temperature $\geq T_{min}$. Euclidean Nariai is $S^2\times S^2$ and the vanishing of the longitudinal circle at the poles of the $S^2$ corresponding to the time and radial coordinates is the infinite temperature limit. $1/T_{min}$ corresponds to the circumference of the great circle.}

The action is
\begin{align}
    I_{EH}+I_{GHY}&=-\pi r_b\left(\frac{4 L^2 r-3 L^2 r_b-4r^3+r_b^3}{L^2-3r_b^2}\right)\nonumber\\
    I_{sub}&=-\frac{r}{T_b}\sqrt{1-\frac{r_b}{r}-\frac{r^2}{L^2}+\frac{r_b^3}{r L^2}}\nonumber\\
    I_{tot}&=I_{EH}+I_{GHY}-I_{sub}
\end{align}
with $r_b$ evaluated on each of the relevant solutions to~(\ref{eq:rbconstraint}). At high temperatures it goes to $-\pi r_b^2$.

The thermodynamic energy and entropy are
\begin{align}
    &E=r\left(1-\sqrt{1-\frac{r_b}{r}-\frac{r^2}{L^2}+\frac{r_b^3}{r L^2}}\right)\nonumber\\
    &S=\pi r_b^2
    \label{eq:ebh}
\end{align}
which apply to both black hole solutions. The first law is satisfied in the usual form by this energy, entropy, and the local temperature at $r$. The energy also satisfies 
\begin{align}
    E-\frac{1}{2}\frac{E^2}{r}=\frac{r^3}{2L^2}+M.
\end{align}
There is no contribution to the entropy from the cosmological horizon because it lies outside the cavity.

We can make various other observations.  First, the total thermodynamic energy is positive. It receives contributions from a gravitational binding energy, the mass parameter of the black hole, and a positive ``vacuum energy" that grows with the volume of the cavity. 
At high temperatures the energy approaches $E=r$, with corrections of order $\sqrt{1-r_b/r}$ ($\sqrt{r_c/r-1}$) for $r<r_N$ ($r>r_N$).

Second, unlike in flat space, the  high temperature limit of the entropy is not $\pi r^2$ for $r>r_N$, because the larger BH solution is  getting smaller as the temperature increases. The cosmological horizon is getting close to $r$ but we do not see its entropy.  The maximum subsystem entropy that can be reached under the constraint of thermal equilibrium is $\pi r_N^2$. The largest area radius is $r\rightarrow L$ and in this case there is no black hole in the equilibrium state.

We can also compute the heat capacities $C=-\beta^2\partial_\beta^2 (\beta F)=TdS/dT$.\footnote{Other thermodynamic aspects of heat capacities in SdS have previously been studied in~\cite{Johnson:2019ayc,Johnson:2019vqf,Dinsmore:2019elr} and the connection with the ``constrained state" proposal of Banks, Fiol, and Morisse~\cite{Banks:2006rx} was noted in~\cite{Dinsmore:2019elr}.} For the empty dS solution, $C=0$ at leading semiclassical order. For the  black hole solutions, we find
\begin{align}
    C=-\frac{4 \pi  r_b^2 \left(L^2-3 r_b^2\right) \left(L^2 (r_b-r)+r^3-r_b^3\right)}{L^4 (3 r_b-2
   r)+2 L^2 \left(r^3-3 r r_b^2-r_b^3\right)+3 r_b^2 \left(2 r^3+r_b^3\right)}.
   \label{bhheatcap}
\end{align}
For both black hole solutions the energy and entropy are  increasing functions of $r_b$. The temperature is an increasing function of $r_b$ for the larger black hole solution and a decreasing function of $r_b$ for the smaller. Therefore the heat capacity is positive for the larger $r_b$ solution and negative for the smaller $r_b$ solution.

\subsection{Cosmological side solution}
We can obtain a fourth solution corresponding to a region of Euclidean SdS with  $\rho\geq r$.  In this case we keep the cosmological horizon and cut out the black hole horizon, taking the conical singularity to be at $r_b$. We refer to this as a ``cosmological side" solution, and it appears more like the outside of a cavity than the inside. However, at the moment we are simply considering the variational problem with a fixed boundary 3-metric, and therefore this solution must be considered on similar footing with the others.

 To have a smooth cosmological horizon we require that the local temperature at the boundary is
$T=T_c/\sqrt{f(r)}$, or
\begin{align}
    T=\frac{3r_c^2-L^2}{4\pi L^2r_c}\frac{1}{\sqrt{1-\frac{r_c}{r}-\frac{r^2}{L^2}+\frac{r_c^3}{r L^2}}}
    \label{eq:rcconstraint}.
\end{align}
The relevant solutions have $r\leq r_c\leq L$. We can make the temperature arbitrarily large for $r<r_N$ by tuning $r_c$ so that $r_b\rightarrow r$, or, for $r>r_N$, by taking $r_c\rightarrow r$. However, there is no high-temperature solution analogous to $r_b\rightarrow 0$, because the thermal circle size is not controlled by the conical deficit at $r_b$. Thus in general there are zero or one relevant solutions for $r_c$.  For any $r$ the minimum temperature is obtained for $r_c\rightarrow L$, and the global minimum temperature is the empty de Sitter temperature with redshift zero, $T=1/2\pi L$, achieved for $r_c\rightarrow L$, $r\rightarrow 0$.

\begin{figure}[t!]
\begin{center}
\includegraphics[width=0.5\linewidth]{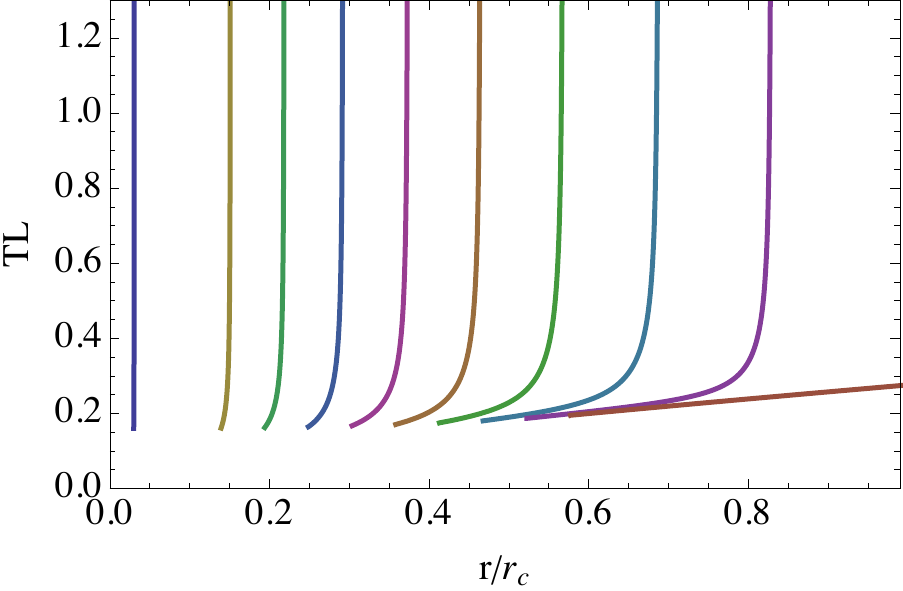}
\caption{
Local temperature at the boundary radius $r$ in units of the dS curvature scale, for a set of $r$ less than the Nariai radius, as a function of cosmological horizon radius and assuming a smooth cosmological horizon. Values of r increase from the left curve to the right curve ($r=r_N$). There are zero or one solutions for a given $T$. }
\label{fig:rcT}
\end{center}
\end{figure}

The various contributions to the action are
\begin{align}
    I_{EH}+I_{GHY}&=\pi r_c\left(\frac{4 L^2 r-3 L^2 r_c-4r^3+r_c^3}{3r_c^2-L^2}\right)\nonumber\\
    I_{sub}&=\frac{r}{T_c}\sqrt{1-\frac{r_c}{r}-\frac{r^2}{L^2}+\frac{r_c^3}{r L^2}}\nonumber\\
    I_{tot}&=I_{EH}+I_{GHY}-I_{sub}.
\end{align}
The thermodynamic energy and entropy of this solution are
\begin{align}
E&=-r\left(1-\sqrt{1-\frac{r_c}{r}-\frac{r^2}{L^2}+\frac{r_c^3}{r L^2}}\right)\nonumber\\
S&=\pi r_c^2.
\label{eq:ecos}
\end{align}
The energy is negative and obeys the relation:
\begin{align}
    E+\frac{1}{2}\frac{E^2}{r}+\frac{r^3}{2L^2}=-M.
\end{align}
 One can also check that $\partial S/\partial E=1/T$, so again the first law is satisfied in the usual form by this energy and temperature, with no additional minus sign like in Eq.~(\ref{SdSfirstlaw}). In the small $r_b, r$ limit we obtain $E=-M$. In other words, we are {\emph{removing}} energy from the ensemble as we increase the black hole mass parameter.

As for the heat capacity, we obtain the same result as Eq.~(\ref{bhheatcap}), but with $r_b$ everywhere replaced by the $r_c$ relevant for the cosmological side solution. (Note that these are not related by Eq.~(\ref{eq:MLeq}) because the $M$ characterizing the cosmological side and black hole side solutions is generally different for a given $T(r)$.) The heat capacity is negative, because the temperature is a decreasing function of $r_c$ while the energy is an increasing function of $r_c$ (even more obviously, the entropy is an increasing function of $r_c$). 

\subsection{Thermodynamics and comparison to the flat space limit}

\begin{figure}[t!]
\begin{center}
\includegraphics[width=0.55\linewidth]{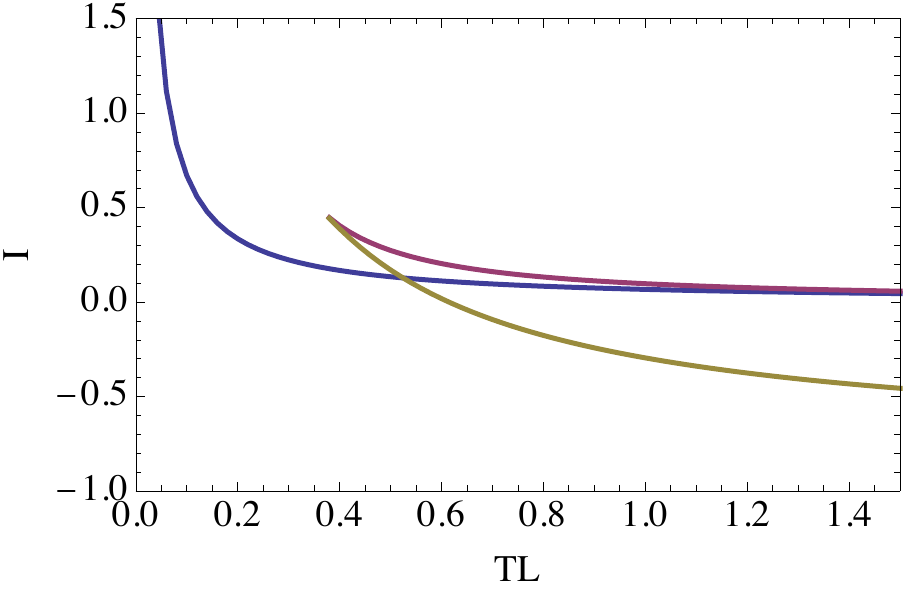}~~
\caption{Euclidean actions of the three black hole-side solutions. In this example $r=L/2$ and the action is shown in units of $L^2/G_N$. At low temperature only the ``empty patch of dS" solution exists. At an intermediate temperature two new black hole solutions appear. At high temperature the lowest action solution is the larger black hole with $r_b\approx r$, and $I_{tot}$ asymptotes to minus the entropy (about -0.8 in these units, off the plot.)}
\label{fig:actionexample}
\end{center}
\end{figure}

The cosmological side solution, when it exists, always has the lowest action  $\beta F = \beta E -S$ of all four solutions, because it is the only solution with negative thermodynamic energy and its entropy is the largest of the four. Therefore  we are compelled to identify it as the semiclassical approximation to the canonical partition function with fixed boundary 3-metric of temperature $T(r)$. However, in the canonical ensemble  $\beta^2\partial_\beta^2 \log Z$ must be nonnegative, because it is equal to the variance of $E/T$. So the cosmological side solution cannot be an approximation to a stable equilibrium state of a canonical ensemble.

Let us briefly recall the zero-c.c.  version of ``gravity in a cavity" studied by York~\cite{York:1986it}. There is no equivalent to the cosmological side solution in asymptotically flat space, and so the problem of the large negative action and negative heat capacity noted above does not arise. The number of solutions, and which one dominates the path integral, are determined by two special values of the dimensionless parameter $p = Tr$, 
$p_1 = 3 \sqrt{3} / (8\pi)  \approx  0.21 $ and $p_2 = 27 / (32 \pi)          \approx  0.27 $.
If $T < p_1/r$, then the only saddle point is $R^3 \times S^1$. This saddle point exists for all temperatures and has zero action at leading order. A negative action corresponding to the free energy of thermal gravitons, $\beta F\sim -VT^3$, is generated by fluctuations. If $p_1/r < T$, there are two other saddle points, corresponding to ``small" and ``large"  black holes. If $T < p_2/r$,  both of these saddles have positive action, and the $R^3 \times S^1$ solution is still the lowest action. If $T > p_2/r$, the action of the large BH is negative and is the lowest, at least at leading order. This is analogous to the Hawking-Page transition~\cite{Hawking:1982dh}.

Many of these qualitative features are shared by the black hole side solutions at positive cosmological constant. If we ignore the cosmological side, we have one, two, or three solutions on the black hole side.  In general, the behavior of the actions mimics the flat space case: at low temperatures, there is only the ``empty region of dS" solution; at intermediate temperatures the black hole solutions appear, but have higher action than empty dS; and finally at higher temperatures the larger black hole solution has the lowest action. This behavior is illustrated in Fig.~\ref{fig:actionexample}.

On the other hand, the existence of the cosmological side solution is a novel feature of the $L>0$ variational problem with fixed boundary 3-metric.  Physically this is not such a surprise that what appears to be the outside of a cavity cannot really be treated as a cavity. It is hard to imagine that we could contain an effective bath inside radii smaller than $r$ that would be necessary to maintain the constraint $T(r)$.

Often negative heat capacity states can be stabilized by considering the microcanonical ensemble. In the next section we discuss the boundary conditions required to convert the canonical to the microcanonical ensemble and the thermodynamic properties.

\section{Modified boundary terms and the microcanonical ensemble}
\label{sec:micro}
We can change the variational problem from fixed temperature to fixed thermodynamic energy by adding a new boundary term to the action.  It is most convenient to express the new term in the ADM formalism, foliating the spacetime into fixed-$t$ hypersurfaces $\Sigma_t$ with normal $N^a$. $\Sigma_t$ intersects the hypersurface at fixed $\rho=r$ on a two dimensional surface $S_t$.  We add to the action a boundary term that takes the form
\begin{align}
    \Delta I= \frac{1}{8\pi} \int dt \int_{S_t} \sqrt{s} N (k - k^0)
    \label{eq:Itotjoined}
\end{align}
if the following gauge condition holds:
\begin{align}
g^{\rho\alpha}=0\;\mbox{at}\;\rho=r\;\mbox{for}\;\alpha\neq\rho.\label{eq:gauge_gra1}
\end{align}
In Eq.~(\ref{eq:Itotjoined}) $s_{ab}$ is the induced metric and $k=s^{ab}\nabla_ar_b$ is the extrinsic curvature on ${S_t}$, where $r^a$ is the unit normal to $S_t$ that is tangent to the constant $t$ surface. The term proportional to $k^0$ is somewhat arbitrary, but in the explicit solutions we consider $S_t$ will be a 2-sphere, and in this case it is convenient to take $k^0$ to be the extrinsic curvature of the surface embedded in flat space. 
The new boundary term in Eq.~(\ref{eq:Itotjoined}) is simply the opposite of an ordinary boundary term in the ADM formalism, and therefore removes this term from the usual ADM action, reducing it to the bulk term alone (cf. Eq.~(\ref{IADM})).

With this modification, the variational problem is well-posed for a different set of boundary conditions on the spatial boundary at $\rho=r$ from the usual Dirichlet conditions. We usually fix the induced metric on the boundary, which amounts to fixing the lapse $N$, shifts $N_a$, and $s_{ab}$. But the modified ADM action has only the bulk term, and its variational problem is well-posed if we fix $k$ instead of $N$ on the boundary (and, as usual, we fix $N_a$ and $s_{ab}$). To see this, first note that the lapse is undifferentiated in the bulk ADM action, so its variation $\delta N$ does not induce a surface term even if $\delta N$ is arbitrary on the boundary. The variation of the boundary term that we removed by adding $\Delta I$ to the usual ADM action cancels out the surface term induced by the variation of the spatial metric in the bulk ADM action. So if $s_{ab}$ is fixed on the boundary, the bulk ADM action must induce a surface term proportional to the integral of $\sqrt{s} N \delta k$ unless $k$ is also fixed on the boundary. (The conditions on $N_a$ are unaffected by our modification.)

The gauge condition \eqref{eq:gauge_gra1} implies that the variation $\delta g^{\rho\alpha}$, $\alpha\neq\rho$, is zero on the boundary. The components $g^{\rho\alpha}$ are not part of the induced metric, so there is no such restriction on their variation in the usual Dirichlet case. It is not restrictive in our case either: since the condition is imposed only on the boundary, the stationarity of the action still provides the full set of field equations in the bulk. Furthermore, for any solution, the gauge condition \eqref{eq:gauge_gra1} can always be satisfied by an appropriate foliation in the neighborhood of the boundary. 

 We consider only spherically symmetric static boundary data, in which $s_{ab}$ is of the form $r^2(d\theta^2+\sin^2\theta d\phi^2)$, $N_a=0$, and $k$ is constant. We assume that the solution with such boundary values is also spherically symmetric and (therefore) static. For such metrics, and working in coordinates where the angular part of the metric takes the usual form $\rho^2(d\theta^2+\sin^2\theta d\phi^2)$, fixing $k$ at $r=\rho$ amounts to fixing $g^{\rho\rho}(r)$. Let us evaluate the new boundary term $\Delta I$ on the black hole and cosmological side cavity solutions. The $S_t$ are two spheres with induced metric
$s_{ab}dx^adx^b = r^2d\Omega^2$, and $r^a=\pm\sqrt{f}\partial_\rho$.  The extrinsic curvatures  are
\begin{align}
k-k^0=\pm\frac{2}{r}\left(\sqrt{f}-1\right),
\end{align}
where $f=g^{\rho\rho}$ is evaluated in each of the respective vacuum solutions at the shell boundary. The subtraction term contributes the $-1$ in the parenthesis; it is not necessary, but we include it here.  So the boundary contribution in the black hole side case is 
\begin{align}
    \Delta I=\frac{4\pi r^2}{8\pi}\beta\left[\frac{2}{r}(\sqrt{f_{in}}-1)\right]
    =\beta r\left[\sqrt{f_{in}}-1\right]=-\beta E,
\end{align}
cf. Eq.~(\ref{eq:ebh}), and in the cosmological side case it is the same, 
\begin{align}
    \Delta I=\frac{4\pi r^2}{8\pi}\beta\left[-\frac{2}{r}(\sqrt{f_{out}}-1)\right]=-\beta E,
\end{align}
cf. Eq.~(\ref{eq:ecos}). 
The new boundary terms are equivalent to adding $-\beta$ times the thermodynamic energy from each region. In other words, they convert the ensemble from canonical to microcanonical, 
\begin{align}
    \log Z\rightarrow S.
\end{align}

At fixed energy and boundary radius, the thermodynamic properties are somewhat different. The  energy of the black hole side solutions is positive and that of the cosmological side solution is negative, so they never need to be considered at the same time. 

Let us discuss the cosmological side first. The negative heat capacity no longer indicates an instability since the energy is fixed. In general there will also be a higher order semiclassical contribution to the thermodynamic quantities from radiation in the cavity emitted by the horizon. However, these corrections can be neglected in first approximation: estimating the radiation energy and entropy as $\calO(r_c^3T_c^4)\lesssim \calO(1/L)$ and $\calO(r_c^3 T_c^3)\lesssim\calO(1)$, both are seen to be minuscule compared to the typical $\calO(L)$ and $\calO(L^2)$ leading order contributions from the classical geometry. Although the radiation heat capacity is positive, once its temperature equilibrates with the horizon, the system is stable. This is a property of microcanonical equilibria: if one subsystem has negative heat capacity, and the other positive, stability requires that the total is negative.

There is a minimum positive energy for which solutions exist, corresponding to the energy of the cavity solution with no horizons. For all larger energies (and fixed boundary radius) there is a single black hole solution which has the minimum classical action. The heat capacity may be positive or negative. The impact of including radiation is again minor for large black holes, but may be more significant for very small ones.  We can estimate the radiation energy and entropy as $\calO(r^3T_b^4)\sim \calO(M (M_p/M)^2(r/r_b)^3)$ and $\calO(r^3 T_b^3)\sim\calO((r/r_b)^3)$. These are subdominant to the classical contributions as long as $r_b/r$ is not too small. If $r_b/r$ is tiny, however, the system will prefer to fill with radiation in order to maximize the entropy.

\section{Removing the boundaries: equilibrium with a membrane}
\label{sec:equilibrium}
Now we consider spacetimes obtained by joining two of the previous solutions at a common boundary at $\rho=r$. We know that unless the resulting spacetime is pure de Sitter there will be some sort of discontinuity at $r$, since there is no completely smooth Euclidean SdS solution.  Let us assume going forward that the time coordinates on each side of the discontinuity are scaled to have the same periodicity, and that the other coordinates are standard spherical coordinates, with cavity boundaries at a common $r$ and in a gauge where the angular part of the metric is always that of the 2-sphere.

 In such coordinates, what sorts of metric discontinuities should we permit?   In this work we focus on configurations that respect local thermodynamic equilibrium. This corresponds to continuity of the temperature at $r$, which is equivalent to extremizing the total entropy at fixed total energy. The discontinuity appears in  $g_{\rho\rho}$. This is equivalent to inserting a thin, rigid, massive membrane at $r$, and it can give rise to a physically viable classical metric if $g_{tt}$ is continuous. 
Continuity of the temperature also implies that the rigid massive membrane is heat conducting. We now explore these metrics in more detail.

A natural interpretation of the cosmological side solution is that it is providing the state of a bath in equilibrium with the black hole side solutions. To realize this we can join the two types of solution at $r$.
Equating the temperatures at $r$, we obtain an equation for $r_c$ characterizing the corresponding cosmological side solution. 
There is a discontinuity in $g_{\rho\rho}$ at $\rho=r$, but the other metric components are continuous. The Hamiltonian boundary terms on either side of the juncture that we added above are now responsible for fixing the total thermodynamic energy in the cavities, and energy can flow from one region to the other.

In fact this procedure of joining two microcanonical cavities is equivalent to considering a closed system, consisting of the gravitational field and a thin shell of matter. This is another reason why the joined system should be in microcanonical equilibrium. We derive in detail the equivalence between two adjacent cavities and a closed system with a physical membrane in Appendix~\ref{appx:membrane}. Here we summarize the argument briefly. We can start by writing the gravitational action of the closed system in a ``radial Hamiltonian" form, where the coordinate $\rho$ plays the role of time. In this form the lapse $N=1/\sqrt{g^{\rho\rho}}$ is undifferentiated in the ADM action and the extension from smooth to discontinuous $N$ is trivial. The gravitational action in this form has no boundary terms of any type.  It can then be rewritten and computed in other convenient forms using the equivalence of various formulations of the action; for example, a Lagrangian form, or a more conventional ``temporal Hamiltonian" form with time coordinate $t$. These other forms generally do have terms localized at $r$ which account for the discontinuities in variables that are differentiated in the action. We use the temporal Hamiltonian for our computations in the appendix since in that form the relevant terms in the action are simple hypersurface terms. We must also add to the full Euclidean action a membrane term,
\begin{align}
I_{\rm matter} =\mu\int_{\So\times\Stw} dt\,d^2\!x\,\sqrt{\gamma}\label{Imatter}.
\end{align}
The variational problem is such that $s_{ab}=s_{ab}^0$, where $s_{ab}^0$ is the standard metric on the 2-sphere, is a constraint  at $\rho=r$, and we impose the gauge condition $g_{\alpha\rho}=0$ for $\alpha\neq \rho$. Otherwise the variations are arbitrary, and in particular the variation of the degree of freedom in $g_{tt}$ gives rise to the new equation
\begin{align}
    k|_{\rho\to r+}+k|_{\rho\to r-}=8\pi\mu.
\end{align}

This equation says that the solutions to the closed system problem are vacuum solutions joined together with a jump in $k$ which is determined by the fixed tension of the rigid membrane. This is equivalent to two cavity solutions in thermal equilibrium with fixed total thermodynamic energy, which is also given by the jump in $k$.\footnote{ More precisely, this is true for (black hole, cosmological horizon) and (empty patch, cosmological horizon) cavity pairs, where the $k^0$ subtraction terms in Eq.~(\ref{eq:Itotjoined}) cancel out from each side. They would not cancel for other pairings, like joining two cavities containing a cosmological horizon. However,  these more exotic configurations are also irrelevant for other reasons. Gluing together two cosmological side solutions requires negative $\mu$, which is unlike the other cases. We can characterize the microcanonical ensemble by $\mu$ and take it to be positive, in which case the double cosmological side solution does not arise. Two  black hole solutions and two empty patch solutions have positive $\mu$, but always have much lower entropy than the mixed combinations and so are irrelevant.} Finally, we find that in both cases the total on-shell action is given by the sum of the horizon entropies, as expected for a closed system in equilibrium: the matter action  cancels against an opposite contribution in the gravitational action, apart from a term equal to the total entropy.

Thus by connecting the two types of cavity solutions, we obtain a solution equivalent to one without boundaries. The energy density and the angular stresses responsible for the rigidity of the shell are singular in the thin-shell limit, but we can imagine they are regularized by a small finite thickness.  Also, we note that the energy conditions and other desirable mechanical stability features of real shells (see e.g.~\cite{Brady:1991np}) may not be satisfied for all values of the parameters; this would be interesting to examine in detail, but we will not pursue it here.


Now let us discuss the thermodynamics of these joined solutions. For the outer region, we take the cosmological side solution. The inner region may contain a  black hole, or  it may be empty dS.\footnote{Other possibilities can be neglected; see previous footnote.} Let us refer to the state of the whole system in the former case as state $A$, and in the latter case as state $B$. The temperature is continuous at $r$ in each state, but in general $T_A(r)\neq T_B(r)$, because we are instead fixing the total energy in the two regions.

It turns out that it is always the case that $S_A<S_B$, so that the black hole solution is never the ground state. This can be seen in the  limiting behaviors or numerically. In state $A$, $T_b$ diverges as $r_b\rightarrow 0$, and so continuity of the temperature requires $f_{out}(A)$, the redshift factor of the cosmological side solution in system $A$, to go to zero as well. Here we have
\begin{align}
    |E_{tot}(A)| &= r |\sqrt{f_{out}(A)}-\sqrt{f_{in}(A)}|\nonumber\\
    &\rightarrow r\sqrt{1-r^2/L^2}
\end{align} 
which is extremal. In state $B$, \begin{align}
|E_{tot}(B)|=r |\sqrt{f_{out}(B)}-\sqrt{1-r^2/L^2}|.
\end{align}
Thus to achieve the same $E_{tot}$ we must also have $f_{out}(B)=0$, and therefore the same $r_c$ as in system $A$. Then the total entropies $S(A)$ and $S(B)$ are equal in the limit that the black hole disappears, but here  the states are the same. Now as we increase $r_b$, $S(A)$ and $S(B)$ both increase, but $S(B)$ increases faster: $f_{out}(B)$ must increase faster than $f_{out}(A)$ to maintain equal values of $E_{tot}$. Finally,  $E_{tot}$ reaches its other extremum at $0$.  In this limit the value of $M$ on the two sides of $r$ is approximately the same. Therefore $S_A\approx \pi r_b^2+\pi r_c^2$, where $r_{b,c}$ are approximately related by~(\ref{eq:MLeq}), and $S_B\approx\pi L^2$. This is just the entropy of an SdS solution compared to that of empty dS with the same cosmological constant, and the latter always has larger entropy. 

\begin{figure}[t!]
\begin{center}
\includegraphics[width=0.5\linewidth]{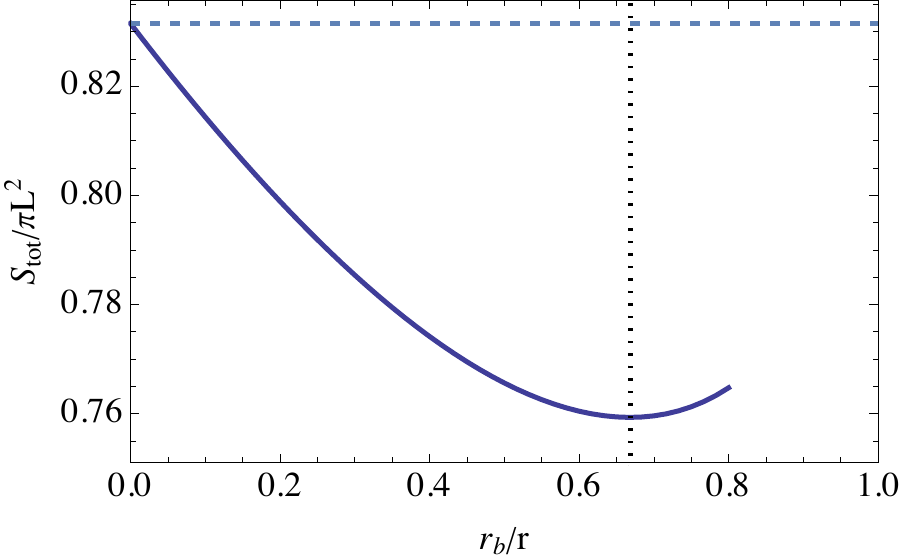}
\caption{The entropy along a family of non-equilibrium states interpolating between equilibrium state $B$ at $r_b=0$, corresponding to an empty patch joined to a cosmological horizon, and equilibrium state $A$ at the minimum of the curve, corresponding to a black hole horizon joined to a cosmological horizon. The total energy and r are fixed to representative values, $E_{tot}=-1/10$ (in units of $L/G_N$) and  $r/L=1/4$.}
\label{fig:Scomparison}
\end{center}
\end{figure}

 A numerical example is shown in Fig.~\ref{fig:Scomparison}. Here we fix $r$ and $E_{tot}$ to representative values and plot the total horizon entropy as a function of $r_b/r$. For general points the system is not in equilibrium and both $g_{tt}$ and $g_{\rho\rho}$ are discontinuous. The equilibrium points are at $r_b=0$, where the entropy is maximal and the solution corresponds to state $B$, and at the minimum of the entropy curve, which corresponds to state $A$.

This is rather unlike what happens in the zero cosmological constant case, where there is no cosmological side solution and the black hole side solutions with boundary admit a canonical ensemble interpretation. There, at high temperatures, a black hole solution has the lowest classical free energy.

 \begin{figure}[t!]
\begin{center}
\includegraphics[width=0.48\linewidth]{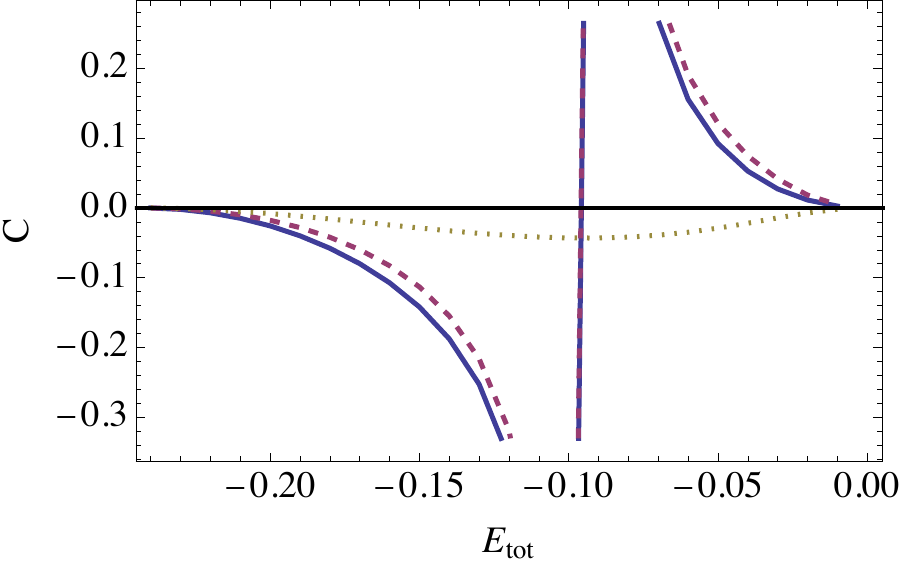}~~
\includegraphics[width=0.48\linewidth]{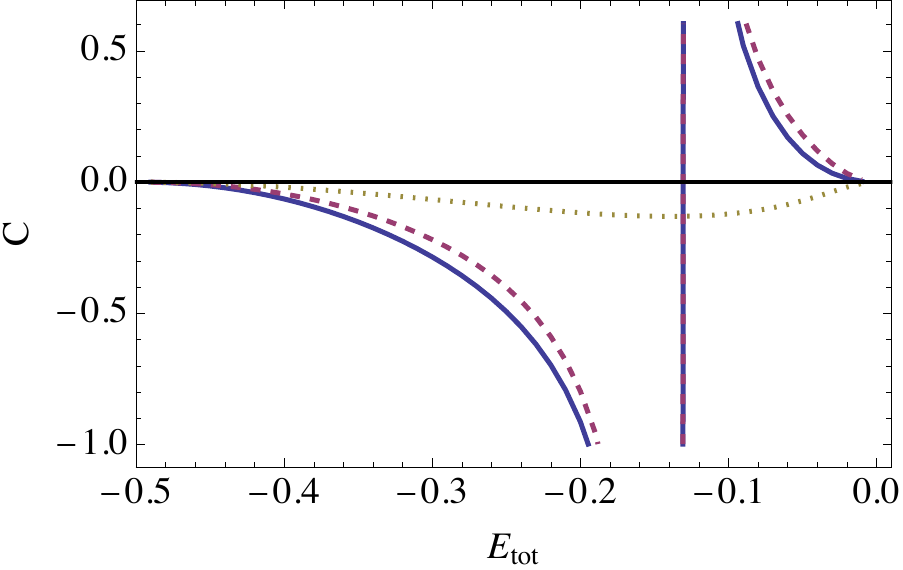}
\caption{Heat capacity of the joined black hole + cosmological horizon solutions as a function of the total thermodynamic energy in the cavities, which is negative and bounded from below. $C$ is in units of $L^2/G_N$ and $E_{tot}$ is in units of $L/G_N$. Lines correspond to the total (solid), black hole side (dashed) and cosmological side (dotted) heat capacities. Although the heat capacity of the black hole side solution changes sign discontinuously, the properties are such that the combined system is always thermodynamically unstable. The two panels correspond to different fixed choices for the juncture location (left: $r/L=1/4$; right: $r/L=3/4$.)}
\label{fig:microheatcap}
\end{center}
\end{figure}

In fact, in the joined ``system+bath" solutions, the equilibrium 
represented by solution 
$A$ is not even metastable. 
In general, when two isolated systems are brought together, stability against small energy fluctuations requires either that both heat capacities are positive, or that the heat capacities are of opposite sign but their sum is {\emph{negative}}. For system $B$, the inner region has vanishing semiclassical heat capacity (a small positive one is generated by quantum fluctuations) and negative  heat capacity in the outer region, so it is thermodynamically stable, at least at this order of approximation. For system $A$, either both sides of the membrane have negative heat capacity, or the black hole side has a large positive value so that the sum is positive. Two examples are plotted in Fig.~\ref{fig:microheatcap}.

Thus we can construct a picture of the evolution of the state $A$ containing the black hole. Fluctuations in one direction lead the black hole to radiate away entirely, leaving the equilibrium state of $B$. Fluctuations in the other direction cause the black hole to grow, and eventually it will consume the shell at $r$. Presumably some energy is liberated in the process and the final state is simply empty de Sitter.

We might further relax the constraints by allowing the wall to move, extremizing the  entropy over $r$. If we fix the total cavity energy and require continuity of $T(r)$, both horizon radii are determined.  It is straightforward to see that the total entropy has a minimum at $r=r_N=L/\sqrt{3}$, with the cosmological side solution approaching the Nariai solution. This is because we can achieve any local temperature $T(r)\geq 1/2\pi L$ on the Nariai solution  by choosing an appropriate circle on the time-radius $S^2$, and at $r_c=r_N$ the cosmological horizon entropy is minimized. 

\section{Summary}
We have analyzed the thermodynamic properties of  spherically symmetric cavities in de Sitter space using Euclidean methods. In the canonical ensemble, we fix a boundary 3-metric with periodic time. 
As in flat space, we find a discrete set of solutions, including cases that contain empty space and cases that contain black holes. Unlike in flat space, there is a solution compatible with the boundary conditions that contains a smooth cosmological horizon. This solution complicates the canonical ensemble interpretation, because it has the lowest action and negative heat capacity. Instead, we convert to the microcanonical ensemble by the addition of appropriate boundary terms to the gravitational action. These boundaries may be isolated, or they may be removed by joining two cavities together. In the latter case metric discontinuities arise, corresponding to physical membranes or out-of-equilibrium states. Here we focused on the membrane case, which permits local thermodynamic equilibrium between a black hole and a cosmological horizon. However, we find that this equilibrium is always unstable, and the stable equilibrium state is an empty shell inside a larger cosmological horizon. This is consistent with the observation that the total horizon entropy of SdS is always less than the cosmological horizon entropy of empty de Sitter with the same cosmological constant.

It would be interesting to extend the analysis here to the microcanonical thermodynamic equilibrium properties of other solutions exhibiting various types of horizons, including other de Sitter black holes and finite causal diamonds with and without black holes~\cite{JV,Banks:2020tox}.

\section*{Acknowledgments}
This work was written with support from the US Department of Energy under grant number DE-SC0015655, and from the DOE Office of High Energy
Physics QuantISED program under an award for the Fermilab Theory Consortium “Intersections of QIS and
Theoretical Particle Physics.”

\appendix
\section{Appendix: Physical Membranes}
\label{appx:membrane}
By joining two cavities with spherically symmetric solutions inside, we can get a spacetime metric that (in Lorentzian signature) resembles the gravitational field of an infinitely thin shell of matter. Indeed, if the cavities are at thermal equilibrium (the proper length of their Euclidean time is the same at the boundary), the metric components are continuous except $g_{\rho\rho}$, which jumps at the juncture. These joint configurations can be thought of as closed systems of the gravitational field and some matter, and we derive this equivalence in detail in this appendix.
\subsection{Action}
To define the Euclidean action of such a closed system, we could start with a nonsingular matter distribution. This approach has the advantage that the action would have no boundary terms. Its disadvantage is that this matter distribution would have more degrees of freedom than what we are ultimately interested in, so the corresponding variational problem would involve more variables than needed for the specification of a spherically symmetric infinitely thin rigid shell. Instead of trying to regularize the matter distribution, we define the Euclidean matter action already in the limit of the thin shell:
\begin{align}
I_{\rm matter} =\mu\int_{\So\times\Stw} dt\,d^2\!x\,\sqrt{\gamma},\label{Imatter}
\end{align}
where $\So$ is the range of the Euclidean time coordinate in a neighborhood of the shell, $\Stw$ is the two-sphere representing the shell in space, and $\gamma_{ab}$ is induced metric on $\So\times\Stw$.

To find the gravitational action in the presence of the shell, we start with the Einstein-Hilbert action of a smooth metric. The gravitational field of the shell has weaker continuity and differentiability properties. To define the action for such metrics, we will rewrite the EH action in a ``radial Hamiltonian" form, in which the extension to metrics of sufficiently weak regularity at the shell is elementary. For this purpose, we introduce the GHY boundary term, with which the  Euclidean Lagrangian action $I_{\calL}=I_{EH}+I_{GHY}$  on a manifold $\cal M$ can be written in the Arnowitt-Deser-Misner (ADM) form as follows:

\begin{align}
\IEH &= -\frac{1}{16\pi}\int_{\cal{M}}d^4x\sqrt{g}(R-2\Lambda)-\frac{1}{8\pi}\int_{\partial\cal M}d^3x\sqrt{\gamma}K=I_{\rm ADM},\label{IE}\\
I_{\rm ADM}&=\frac{1}{16\pi }\int_{\tau_1}^{\tau_2} d\tau\int_{\Sigma_\tau}d^3x\,\left(\pi^{ab}\partial_\tau h_{ab}-N{\cal H}-N_a{\cal H}^a\right)+B.\label{IADM}
\end{align}
Here $\Sigma_\tau$ is a foliation of $\cal M$, $h_{ab}$ is the induced metric on $\Sigma_\tau$, and $\partial_\tau h_{ab}$ is its $\tau$-derivative ($\partial_\tau h_{ab}=h_a^ch_b^d{\cal L}_{\tau}h_{cd}$, where ${\cal L}_\tau$ is the Lie-derivative with respect to $\tau^a$ defined by $\tau^a\nabla_a\tau=1$.) $K=\gamma^{ab}\nabla_a n_b$, where $n^a$ is the outward unit normal to the boundary and $\gamma_{ab}$ is the induced metric on it. $N$ and $N_a$ are the lapse function and shift vectors ($N=1/\sqrt{g^{\tau\tau}}$, $N_i=N^2g_{ij}g^{j\tau}$, where $i$ and $j$ label coordinates other than $\tau$), $\cal H$ and ${\cal H}^a$ are the usual Hamiltonian and momentum constraints, and $\pi^{ab}$ is the conjugate momentum:
\begin{align}
    \pi^{ab}&=\frac{\sqrt{h}}{2N}G^{abcd}(\partial_\tau h_{ab}-D_aN_b-D_bN_a),\label{momentum}\\
    G^{abcd}&=\frac{1}{2}(h^{ac}h^{bd}+h^{ad}h^{bc})-h^{ab}h^{cd},\nonumber
\end{align}
where $D_a$ is the covariant derivative on $\Sigma_\tau$ compatible with $h_{ab}$. The constraints are functionals of the canonical variables $h_{ab}$ and $\pi^{ab}$.\footnote{ They have the same form as in Lorentzian signature, except that the spatial curvature scalar appears with a positive sign in the Hamiltonian constraint.} In addition to $\Sigma_{\tau_{1,2}}$, $\partial{\cal M}$ may contain an additional (not necessarily connected) boundary $\cal T$, in which case there is a boundary term $B$ (an integral of an expression of the metric and its first derivatives on $\cal T$). Actually, in our applications, when $\cal T$ is present, $\tau$ is periodic, so $\partial\cal M=\cal T$, and when $\tau$ is not periodic, $\Sigma_\tau$ has no boundary, so $\partial{\cal M}=\Sigma_{\tau_1}\cup\Sigma_{\tau_2}$.  In the gauge 
\begin{align}
g^{\rho\alpha}=0\;\mbox{at}\;\rho=r\;\mbox{for}\;\alpha\neq\rho,\label{eq:gauge_gra}
\end{align}
$B$ takes the following form:

\begin{align}
B=-\frac{1}{8\pi}\int_{\tau_1}^{\tau_2}d\tau\int_{S_\tau}d^2x\sqrt{s}\left(Nk+N_a\frac{\pi^{ab}}{\sqrt{h}}n_b\right),\label{B}   
\end{align}
where $S_\tau$ is the (not necessarily connected) intersection of $\Sigma_\tau$ and $\cal T$, and $k=s^{ab}\nabla_an_b$, where $s_{ab}$ is the induced metric on $S_\tau$.

\begin{figure}[t!]
\begin{center}
\includegraphics[width=0.7\linewidth]{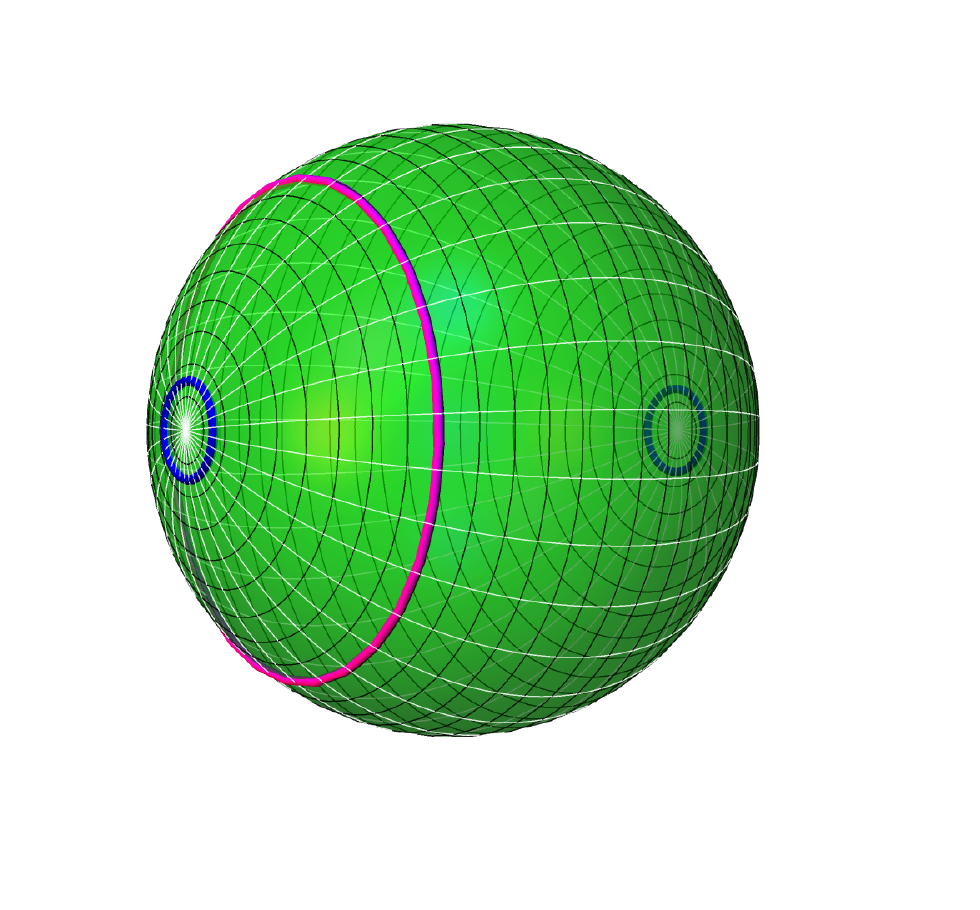}
\caption{
A cartoon of the slicings and boundaries. White mesh lines represent curves of constant $t$, corresponding to the temporal Hamiltonian foliation, and black mesh lines represent curves of constant $\rho$, corresponding to the radial Hamiltonian foliation. The magenta hoop denotes the juncture at $\rho=r$ (shown smooth for simplicity, but in general there is a metric discontinuity on this surface.) Other angular directions are suppressed and the complete juncture is  $\So\times\Stw$. The small blue hoops represent fiducial boundaries at $\rho=r_{1,2}$, which are convenient to introduce in the Hamiltonian formalism. The regions $\rho<r_1$ and $\rho>r_2$ are handled in the Lagrangian formalism.}
\label{fig:sphere}
\end{center}
\end{figure}

Let $\cal M$ be a neighborhood of the shell with a radial coordinate $\rho$, $r_1\leq\rho\leq r_2$, such that the constant $\rho$ surfaces are homeomorphic to $\So\times\Stw$ and the shell is located at $\rho=r$. The boundaries at $r_{1,2}$ are introduced for convenience in the Hamiltonian formalism, but they are only artifacts and we will eventually handle the regions $\rho<r_1$ and $\rho>r_2$ in the Lagrangian formalism. If we choose $\tau=\rho$ in \eqref{IADM}, we obtain what we may call the radial Hamiltonian form of $\IEH$, in which the radius plays the role the time variable. A more conventional choice is when $\tau$ is the Euclidean time $t$ and corresponds to what we will call the temporal Hamiltonian form. These choices will be indicated by the superscript in $\Irad$ and $\Itemp$. In the radial form, $\partial\cal M$ consists of $\Sigma_{\tau_{1,2}}$, which are the $\rho=r_{1,2}$ surfaces, so there is no additional boundary $\cal T$ and $B=0$. Since $N$ is undifferentiated and occurs only in the four-dimensional spacetime integral in \eqref{IADM}, the extension from smooth variables $(N, N_a, h_{ab})$ to configurations with discontinuous $N$ (and $N_a$) is straightforward because $\Irad$ is already meaningful for such variables. On the other hand, the advantage of the temporal Hamiltonian form is that the contributions to the on-shell action are all localized on hypersurfaces. A cartoon of the important radial coordinates and the slicings is shown in Fig.~\ref{fig:sphere}.

 We can use the radial Hamiltonian formulation to easily construct the action in the region $\calM$,  then translate it into an equivalent temporal Hamiltonian form to derive the equations of motion and on-shell value of the action on $\calM$.  The relationship is simply:
\begin{align}
\Irad(r_1\leq\rho\leq r_2) &= \Irad(r_1\leq\rho\leq r) + \Irad(r\leq\rho\leq r_2)\nonumber\\
&= \IEH(r_1\leq\rho\leq r) + \IEH(r\leq\rho\leq r_2)\label{grav_shell}\\
&= \Itemp(r_1\leq\rho\leq r) + \Itemp(r\leq\rho\leq r_2).\nonumber
\end{align}
The decomposition into two regions  in the first line is achieved  by splitting the $\tau$-integral in \eqref{IADM}. There are no boundary terms at this stage and the radial Hamiltonian action is entirely bulk.   The second line is obtained by \eqref{IE} applied to the respective region, and the third line is the result of expressing $\IEH$ in the ADM form, this time using the Euclidean coordinate $t$ as $\tau$. Unlike $\Irad$, the other two forms $\IEH$ and $\Itemp$ have boundary terms at $\rho=r_{1,2}$ and also at the shell ($\rho=r$). Ultimately we will extend $\calM$ to include contributions to the action from $\rho<r_1$ and $\rho> r_2$; this is most easily done in the Lagrangian formalism and will be described  below.

The full action in the region $r_1\leq\rho\leq r_2$ in temporal Hamiltonian form is:
\begin{align}
    I_{\calM} = \Itemp(r_1\leq\rho\leq r) + \Itemp(r\leq\rho\leq r_2) + I_{\rm matter}.\label{full_shell}
\end{align}

\subsection{Field Equations}
The field equations are obtained by varying $g_{ab}$, or equivalently, $(N, N_a, h_{ab})$ of the Hamiltonian form we chose, with induced metric on the surfaces $\rho=r_{1,2}$ held fixed. We assume that the shell is rigid and spherically symmetric, so the induced metric $s_{ab}$ on the intersections $S_t$ of the Euclidean time slices and $\rho=r$ surface is the standard metric $s^0_{ab}$ on the two-sphere of radius $r$. The expression \eqref{B} for the boundary term $B$ is valid if $g_{\rho\alpha}=0$, $\alpha\neq\rho$, at $\rho=r$.
This gauge condition makes it possible to express the determinant of the metric on $\So\times\Stw$ in \eqref{Imatter} in terms of the lapse function $N$ of the temporal Hamiltonian form as
\begin{align}
\sqrt{\gamma}=N\sqrt{s},
\end{align}
which was also used in the derivation of \eqref{B}.\footnote{
We get the full set of field equations even if this gauge condition is imposed before varying the action. This is because the only equation we might miss is an equation relating fields on the shell coming from the variation of $g_{\rho\alpha}$ at $\rho=r$. There is no boundary term arising from the variation of $g_{\rho\alpha}$ in the gravitational action because the surface terms generated by integrating by parts in the derivation of the Euler-Lagrange equations are cancelled out by the variation of the boundary term as long as the induced metric $g_{\alpha\beta}$, $\alpha,\beta\neq\rho$, is kept constant at the boundary. It is this property that we achieve by including the GHY term in the Lagrangian form of the action. In the radial Hamiltonian form, there is no boundary term because $g_{\rho\alpha}$, $\alpha\neq\rho$, are the components of the shift vector, which is undifferentiated in the bulk term, so its variation does not produce boundary terms.}

We  derive the field equations using the temporal Hamiltonian form of the  action, using 
$(N,N_a,h_{ab},\pi^{ab})$ 
as our field variables. $N$ and $N_a$ are fixed at $\rho=r_{1,2}$. At $\rho=r$, the metric $h_{ab}$ is subject to the constraint $s^{\phantom{0}}_{ab}=s^0_{ab}$ and satisfies the gauge condition $g_{\rho\alpha}=0$, $\alpha\neq\rho$. Otherwise the variations are arbitrary. If $N$ and $N_a$ were also fixed at $\rho=r$, the induced metric would be fixed on the boundary in $\Itemp(r_1\leq\rho\leq r)$ and $\Itemp(r\leq\rho\leq r_2)$, so the variation would not produce any boundary term and we would get only the vacuum equations in the region $\rho\neq r$. The only additional equations arise from the variation of $N$ and $N_a$ in $I_{\rm matter}$ and the boundary terms $B$ at $\rho=r$:
\begin{align}
k|_{\rho\to r+}+k|_{\rho\to r-}=8\pi\mu,\label{kS_jump}\\
\left.\frac{\pi^{ab}n_b}{\sqrt{h}}\right\vert_{\rho\to r+}+\left.\frac{\pi^{ab}n_b}{\sqrt{h}}\right\vert_{\rho\to r-}=0.\label{pi_jump}
\end{align}
 The integral of $k$ on $S_t$ is the derivative of the area of $S_t$ as each of its points is moved an equal distance along the outward unit normal. Therefore Eq.~(\ref{kS_jump}) relates $\mu$ to the discontinuity in the rate at which the area of $S_t$ is changing as  its points are moved continuously along directions normal to the shell.

When we join two SdS cavity solutions with the same boundary temperature $T(r)$, using the boundary term prescription of Sec.~\ref{sec:micro} at the juncture, the combined geometry is a solution to a variational problem where the jump in $k$ is fixed while the boundary temperature can vary. What we have shown here is that (in the relevant cases, where the $k^0$ subtraction term in Eq.~(\ref{eq:Itotjoined}) cancels between the two cavities), this is equivalent to considering a closed system with a thin, rigid, massive membrane.

\subsection{Total On-shell Action}
Now we compute the total action on-shell, again working in the temporal Hamiltonian formalism, where all of the contributions to the on-shell action take the form of integrals over hypersurfaces.

The variation of the lapse and shift on the $\rho=r$ surface produce Eqs.~(\ref{kS_jump}) and~(\ref{pi_jump}), but the action is linear in $N$ and $N_a$, so its terms are  proportional to these equations.

In other words, when evaluated on a solution, the matter action and the boundary terms of the gravitational action at $\rho=r$ completely cancel out and there is no contribution from the shell. Furthermore, if the solution is stationary with respect to the Euclidean time ($\partial_t h_{ab}=0$, as in the cases we consider),  the action reduces to the boundary terms $B$ at $\rho=r_{1,2}$, plus contributions from $\rho<r_1$ and $\rho>r_2$. 

 For the field configurations we are interested in, the  boundaries at $\rho=r_{1,2}$  are introduced only because the coordinates of the Hamiltonian form have coordinate singularities at $\rho=0$ or the horizons (see Fig.~\ref{fig:sphere}.) The solutions themselves extend smoothly all the way to zero or the horizons. We can take $r_{1,2}$ infinitesimally close to these points. Since $N_a=0$ and $\pi^{ab}=0$, the second term in \eqref{B} is zero. The first term also vanishes in the limit $r_1\to r_h$ because $N\to 0$ and $k$ remains bounded. The same applies to the other side of the shell. Thus, all that remains of the action of smooth solutions is  contributions  from the regions beyond $r_{1,2}$. We compute these contributions, thus obtaining the total action, using the Lagrangian form of the action for $\rho<r_1$ and $\rho>r_{2}$. This technique is due to~\cite{Banados:1993qp,Teitelboim:2001skl}. 
We enclose the neighborhood of the horizon in the region $\rho<r$ with a sphere at $\rho=r_1$, and 
we make a similar excision on the other side of the shell at $\rho=r_2$. 
In the limit that the neighborhoods that shrink to the horizons, the bulk Einstein-Hilbert action vanishes and only the GHY terms contribute. The GHY term at $r_1$ points toward larger $r$ and the term at $r_2$ toward smaller $r$. The result in the limit is
\begin{align}
    I_{tot} = \sum_{r_1,r_2} I_{GHY} = -\sum A_h/4
\end{align}
 where the sum is over the horizons and $A_h$ is the horizon area.\footnote{This result is valid even in the absence of horizons. In this case, the area radius is zero at the excised point and even though $N$ does not go to zero as we approach it, $\int_{S_t}d^2x\sqrt{s}k$ does, so the limit of $B$ is still zero. Likewise the limit of the GHY term of the neighborhood surrounding the excision goes to zero if there is no horizon.}

\bibliography{euclideands_refs}{}

\providecommand{\href}[2]{#2}\begingroup\raggedright\begin{thebibliography}{10}

\bibitem{Banks:2000fe}
T.~Banks, ``{Cosmological breaking of supersymmetry?},''
  \href{http://dx.doi.org/10.1142/S0217751X01003998}{{\em Int. J. Mod. Phys. A}
  {\bfseries 16} (2001) 910--921},
  \href{http://arxiv.org/abs/hep-th/0007146}{{\ttfamily arXiv:hep-th/0007146}}.

\bibitem{Banks:2006rx}
T.~Banks, B.~Fiol, and A.~Morisse, ``{Towards a quantum theory of de Sitter
  space},'' \href{http://dx.doi.org/10.1088/1126-6708/2006/12/004}{{\em JHEP}
  {\bfseries 12} (2006) 004},
  \href{http://arxiv.org/abs/hep-th/0609062}{{\ttfamily arXiv:hep-th/0609062}}.

\bibitem{Banks:2013fr}
T.~Banks and W.~Fischler, ``{Holographic Theory of Accelerated Observers, the
  S-matrix, and the Emergence of Effective Field Theory},''
  \href{http://arxiv.org/abs/1301.5924}{{\ttfamily arXiv:1301.5924 [hep-th]}}.

\bibitem{Banks:2020zcr}
T.~Banks and W.~Fischler, ``{Holographic Space-time, Newton`s Law, and the
  Dynamics of Horizons},'' \href{http://arxiv.org/abs/2003.03637}{{\ttfamily
  arXiv:2003.03637 [hep-th]}}.

\bibitem{Susskind:2021dfc}
L.~Susskind, ``{Black Holes Hint Towards De Sitter-Matrix Theory},''
  \href{http://arxiv.org/abs/2109.01322}{{\ttfamily arXiv:2109.01322
  [hep-th]}}.

\bibitem{Ginsparg:1982rs}
P.~H. Ginsparg and M.~J. Perry, ``{Semiclassical Perdurance of de Sitter
  Space},'' \href{http://dx.doi.org/10.1016/0550-3213(83)90636-3}{{\em Nucl.
  Phys. B} {\bfseries 222} (1983) 245--268}.

\bibitem{Gross:1982cv}
D.~J. Gross, M.~J. Perry, and L.~G. Yaffe, ``{Instability of Flat Space at
  Finite Temperature},'' \href{http://dx.doi.org/10.1103/PhysRevD.25.330}{{\em
  Phys. Rev. D} {\bfseries 25} (1982) 330--355}.

\bibitem{York:1986it}
J.~W. York, Jr., ``{Black hole thermodynamics and the Euclidean Einstein
  action},'' \href{http://dx.doi.org/10.1103/PhysRevD.33.2092}{{\em Phys. Rev.
  D} {\bfseries 33} (1986) 2092--2099}.

\bibitem{DFeuclidean2}
P.~Draper and S.~Farkas, ``{de Sitter Black Holes as Constrained States\\ in
  the Euclidean Path Integral},'' {\em (to appear)} .

\bibitem{svesko2022quasilocal}
A.~Svesko, E.~Verheijden, E.~P. Verlinde, and M.~R. Visser, ``Quasi-local
  energy and microcanonical entropy in two-dimensional nearly de sitter
  gravity,'' 2022.

\bibitem{Johnson:2019ayc}
C.~V. Johnson, ``{de Sitter Black Holes, Schottky Peaks, and Continuous Heat
  Engines},''
\href{http://arxiv.org/abs/1907.05883}{{\ttfamily arXiv:1907.05883 [hep-th]}}.

\bibitem{Johnson:2019vqf}
C.~V. Johnson, ``{Specific Heats and Schottky Peaks for Black Holes in Extended
  Thermodynamics},''
\href{http://arxiv.org/abs/1905.00539}{{\ttfamily arXiv:1905.00539 [hep-th]}}.

\bibitem{Dinsmore:2019elr}
J.~Dinsmore, P.~Draper, D.~Kastor, Y.~Qiu, and J.~Traschen, ``{Schottky Anomaly
  of deSitter Black Holes},''
\href{http://arxiv.org/abs/1907.00248}{{\ttfamily arXiv:1907.00248 [hep-th]}}.

\bibitem{Hawking:1982dh}
S.~W. Hawking and D.~N. Page, ``{Thermodynamics of Black Holes in anti-De
  Sitter Space},'' \href{http://dx.doi.org/10.1007/BF01208266}{{\em Commun.
  Math. Phys.} {\bfseries 87} (1983) 577}.

\bibitem{Brady:1991np}
P.~R. Brady, J.~Louko, and E.~Poisson, ``{Stability of a shell around a black
  hole},'' \href{http://dx.doi.org/10.1103/PhysRevD.44.1891}{{\em Phys. Rev. D}
  {\bfseries 44} (1991) 1891--1894}.

\bibitem{JV}
T.~Jacobson and M.~Visser, ``{Gravitational Thermodynamics of Causal Diamonds
  in (A)dS},'' \href{http://dx.doi.org/10.21468/SciPostPhys.7.6.079}{{\em
  SciPost Phys.} {\bfseries 7} no.~6, (2019) 079},
  \href{http://arxiv.org/abs/1812.01596}{{\ttfamily arXiv:1812.01596
  [hep-th]}}.

\bibitem{Banks:2020tox}
T.~Banks, P.~Draper, and S.~Farkas, ``{Path Integrals for Causal Diamonds and
  the Covariant Entropy Principle},''
  \href{http://dx.doi.org/10.1103/PhysRevD.103.106022}{{\em Phys. Rev. D}
  {\bfseries 103} no.~10, (2021) 106022},
  \href{http://arxiv.org/abs/2008.03449}{{\ttfamily arXiv:2008.03449
  [hep-th]}}.

\bibitem{Banados:1993qp}
M.~Banados, C.~Teitelboim, and J.~Zanelli, ``{Black hole entropy and the
  dimensional continuation of the Gauss-Bonnet theorem},''
  \href{http://dx.doi.org/10.1103/PhysRevLett.72.957}{{\em Phys. Rev. Lett.}
  {\bfseries 72} (1994) 957--960},
  \href{http://arxiv.org/abs/gr-qc/9309026}{{\ttfamily arXiv:gr-qc/9309026}}.

\bibitem{Teitelboim:2001skl}
C.~Teitelboim, ``{Gravitational thermodynamics of Schwarzschild-de Sitter
  space},'' in {\em {Meeting on Strings and Gravity: Tying the Forces
  Together}}, pp.~291--299.
\newblock 2001.
\newblock \href{http://arxiv.org/abs/hep-th/0203258}{{\ttfamily
  arXiv:hep-th/0203258}}.

\end{thebibliography}\endgroup
\bibliographystyle{utphys}

\end{document}